\documentclass[sigconf]{acmart}

\usepackage{multirow}
\usepackage{arydshln}
\usepackage{booktabs}
\usepackage{amsmath}
\usepackage{makecell}
\usepackage[table,dvipsnames]{xcolor}
\usepackage{enumitem}
\usepackage[ruled,vlined]{algorithm2e}
\usepackage[normalem]{ulem}
\usepackage{graphicx}
\usepackage{eso-pic}

\renewcommand\footnotetextcopyrightpermission[1]{}
\AtBeginDocument{%
  }

\setcopyright{acmlicensed}
\copyrightyear{2018}
\acmYear{2018}
\acmDOI{XXXXXXX.XXXXXXX}

\newcommand{\LongCatHeader}{%
  \AddToShipoutPictureFG*{%
    \AtPageUpperLeft{%
      \raisebox{-1.25cm}[0pt][0pt]{%
        \hspace{1.9cm}%
        \begin{minipage}{\dimexpr\paperwidth-3.8cm\relax}
        \includegraphics[width=0.15\textwidth]{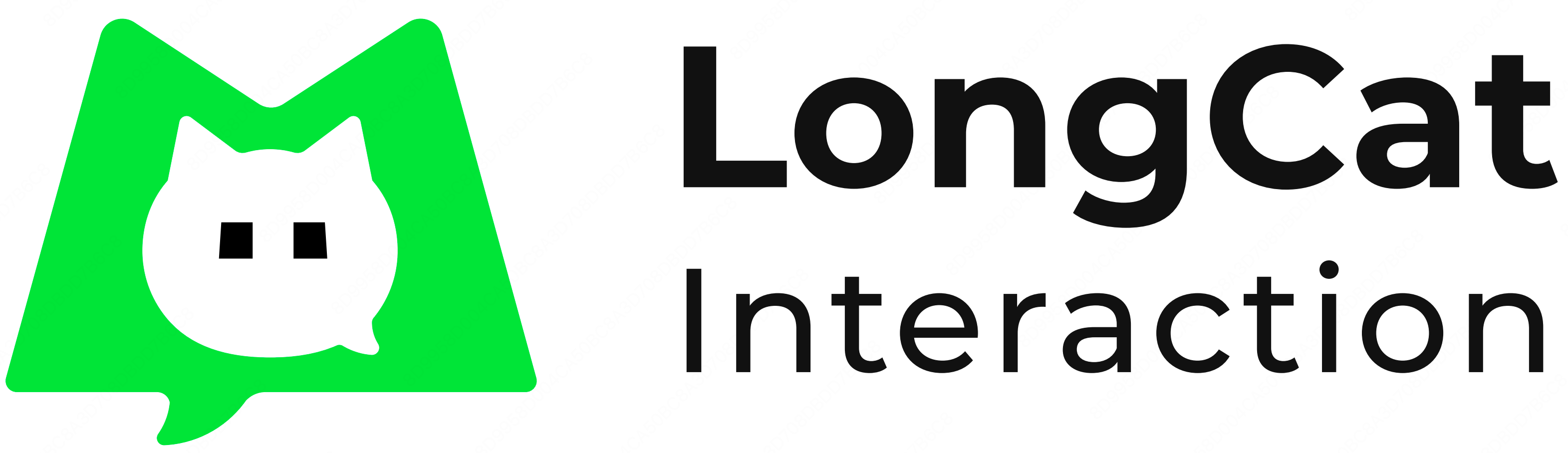}\par
          \rule{\linewidth}{0.3pt}
        \end{minipage}%
      }%
    }%
  }%
}

\title{Factorized Latent Reasoning for LLM-based Recommendation}

\renewcommand{\shortauthors}{Gao et al.}

\author{Tianqi Gao}
\authornote{These authors contributed equally to this work.}
\affiliation{%
  \institution{Independent Researcher}
  \country{China}}
\email{tianqig358@gmail.com}

\author{Chengkai Huang}
\authornotemark[1]
\affiliation{%
  \institution{Macquarie University}
  \institution{University of New South Wales}
  \city{Sydney}
  \country{Australia}}
\email{chengkai.huang1@unsw.edu.au}

\author{Zihan Wang}
\affiliation{%
 \institution{Meituan LongCat Interaction Team}
 \city{Beijing}
 \country{China}}
\email{wangzihan14@meituan.com}
\authornote{Corresponding author.}

\author{Cao Liu}
\affiliation{%
 \institution{Meituan LongCat Interaction Team}
 \city{Beijing}
 \country{China}}
\email{liucao@meituan.com}
 
\author{Ke Zeng}
\affiliation{%
 \institution{Meituan LongCat Interaction Team}
 \city{Beijing}
 \country{China}}
\email{zengke02@meituan.com}

\author{Lina Yao}
\affiliation{%
  \institution{University of New South Wales}
  \city{Sydney}
  \state{NSW}
  \country{Australia}}
\email{lina.yao@unsw.edu.au}

\begin{abstract}
Large language models  (LLMs) have recently been adopted for recommendation by framing user preference modeling as a language generation problem. However, existing latent reasoning approaches typically represent user intent with a single latent vector, which struggles to capture the inherently multi-faceted nature of user preferences.
We propose Factorized Latent Reasoning (FLR), a novel framework for LLM-based sequential recommendation that decomposes latent reasoning into multiple disentangled preference factors. FLR introduces a lightweight multi-factor attention module that iteratively refines a latent thought representation, where each factor attends to distinct aspects of the user’s interaction history. To encourage diversity and specialization, we design orthogonality, attention diversity, and sparsity regularization objectives, and dynamically aggregate factor contributions for the final prediction.
We further integrate FLR with an efficient reinforcement learning strategy based on group-relative policy optimization, enabling stable alignment directly in the latent reasoning space. 
Experiments on multiple benchmarks show that FLR consistently outperforms strong baselines while improving robustness and interpretability. Our data and code are available at \uline{\url{https://github.com/ToAdventure/FLR}}.
\end{abstract}



\begin{document}

\LongCatHeader
\maketitle

\vspace{-0.9em}

\section{Introduction}

\begin{figure*}[t]
    \centering
    {
        \includegraphics[width=\linewidth]{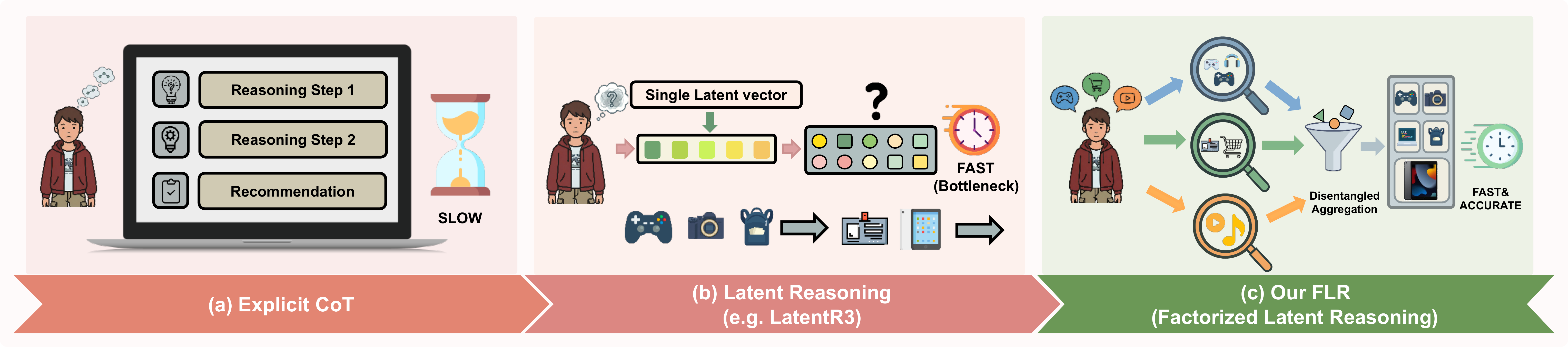}
    }
    \vspace{-1em}
    \caption{Schematic comparison of three reasoning paradigms: (a) Explicit CoT Reasoning, (b) Latent Reasoning, and (c) our Factorized Latent Reasoning (FLR).} 
    \label{fig:intro}
    \vspace{-1em}
\end{figure*}

Recommender systems play a central role in personalized information filtering across domains such as E-commerce, video streaming, and music services. They aim to predict user preferences based on historical interactions, enabling systems to suggest items that are likely to be relevant to users’ interests \cite{huang2025listwise,huang2026generative,huang2026dual}. Sequential recommendation, in particular, leverages user behavior sequences to better capture evolving preferences over time \cite{huang2023dual, huang2023modeling,ye2025beyond,ye2026gaussian}. Traditional approaches in this area typically rely on simple forward encoding of interaction histories, predicting the next item from the final hidden state representation of the sequence encoder. However, such direct computation paradigms often struggle to model the complex and nuanced dynamics of user behavior in practice \cite{gennext_sigir25, DBLP:journals/fcsc/PanPWYM26,huang2025towards,huang2025survey}.

Recent advances in large language models (LLMs) have opened up new opportunities to incorporate deeper reasoning capabilities into recommender systems \cite{huang2025survey}. Early attempts have used explicit reasoning techniques such as chain-of-thought (CoT) prompting or multi-step textual deliberation to break down the recommendation task into intermediate reasoning steps \cite{DBLP:journals/corr/abs-2502-02061}. While this explicit reasoning paradigm can improve semantic understanding, it has two major drawbacks: the need for high-quality CoT annotations and substantial inference overhead due to autoregressive token generation, which makes it difficult to scale to large item vocabularies and latency-sensitive environments.

To overcome these limitations, researchers have begun exploring latent reasoning frameworks that perform reasoning in a continuous representation space instead of generating explicit tokens. This paradigm allows models to iteratively refine internal representations before making recommendations, achieving a trade-off between reasoning capacity and inference cost \cite{DBLP:journals/corr/abs-2505-16865,zhang2025reinforced,huang2025embedding}. For instance, LARES introduces depth-recurrent latent reasoning to sequential recommendation, enabling iterative refinement of latent user representations without relying on discrete reasoning tokens \cite{DBLP:journals/corr/abs-2505-16865}. Similarly, frameworks like Reinforced Latent Reasoning optimize latent reasoning through reinforcement learning, further reducing dependency on explicit reasoning supervision \cite{zhang2025reinforced,liu2021semi}.

Despite these advances, existing latent reasoning approaches primarily compress user preferences into a single latent vector or token sequence at each reasoning step. This simplification inherently assumes that complex user intent can be captured by a monolithic representation, which may obscure fine-grained and diverse user motivations. From a broader perspective, recommender systems have long relied on latent factor models to capture multi-faceted user preferences in collaborative filtering, where each latent factor represents a distinct dimension of user–item interaction \cite{matrix_factorization_survey, 9720218}. Empirically, increasing the number of latent factors in matrix factorization improves recommendation quality until an optimal balance between expressiveness and generalization is reached, highlighting the importance of capturing multiple underlying preference components.
In real-world settings, user preferences are inherently multifaceted and heterogeneous. Modeling such complex intent with a single latent reasoning vector risks information entanglement and loss of explanatory power. Prior work in social recommendation has demonstrated the value of disentangling latent relations into multiple facets to improve representation quality and recommendation performance \cite{9720218}.

In this paper, we argue that effective latent reasoning for recommendation must be factorized into multiple disentangled components, each capturing a distinct aspect of user intent. We present Factorized Latent Reasoning (FLR), a novel framework for sequential recommendation that decomposes the reasoning process into multiple latent factors. Instead of conditioning recommendations on a single reasoning vector, FLR maintains a set of disentangled latent factors that iteratively attend to different semantic facets of the interaction history. These factors are dynamically aggregated to form a comprehensive user representation for final item prediction.

To encourage meaningful specialization and reduce redundancy among these latent factors, we introduce several regularization objectives, such as orthogonality constraints, attention-diversity regularization, and sparsity-guided factor updates. Operating entirely within latent space, FLR avoids the pitfalls of explicit chain-of-thought generation while capturing richer multi-aspect preference structure. Beyond representation learning, we integrate FLR with an efficient reinforcement learning strategy inspired by recent latent reasoning frameworks. Instead of sampling reasoning paths token by token, our approach performs latent exploration via Group Relative Policy Optimization (GRPO) in the factorized latent space, enabling stable alignment of latent reasoning factors with the recommendation objective.


In this paper, our main contributions are:
\begin{itemize}
    \item We propose \emph{Factorized Latent Reasoning} (FLR) for LLM-based recommendation, decomposing user preference into $K$ disentangled latent reasoning factors.
    \item We design a factor-wise attention and gating module that dynamically aggregates disentangled latent preference factors to form the final latent thought representation for recommendation.
    \item We propose disentanglement-oriented objectives (orthogonality, attention diversity, and sparse routing) to ensure non-redundant and interpretable factors.
    \item We integrate FLR with GRPO, enabling reinforcement learning in the factorized latent reasoning space.
\end{itemize}

\section{Related Work}

\subsection{LLM-based Recommendation}
Recent studies have explored the integration of LLMs into recommender systems, particularly through fine-tuning paradigms that align LLMs with task-specific objectives. A major research direction focuses on incorporating reasoning signals into recommendation models, often by making the decision process explicit. EXP3RT~\cite{DBLP:conf/sigir/Kim0CKCY025}, ReasoningRec~\cite{DBLP:conf/naacl/BismayDC25}, SLIM~\cite{DBLP:conf/www/WangTHYLZZPW24}, and Rec-SAVER~\cite{tsai-etal-2024-leveraging} leverage large models to generate Chain-of-Thought (CoT) supervision, which is then distilled or transferred to smaller recommendation models. 
Another line of work seeks to externalize user preferences by transforming implicit behavioral signals into explicit natural language descriptions. CoT4Rec~\cite{DBLP:conf/aaai/YueYZSLW25} and Reason4Rec~\cite{DBLP:journals/corr/abs-2502-02061} adopt clustering strategies or review-based proxies to construct textual preference representations or reasoning traces that guide model training. Complementarily, CoT-Rec~\cite{DBLP:conf/sigir/LiuY0ZGZLSG25} and R2Rec~\cite{DBLP:journals/corr/abs-2506-05069} integrate reasoning modules into the retrieval and ranking pipeline, enabling intermediate analytical steps such as preference decomposition or interaction-level reasoning.
More recently, several works have explored extending reasoning depth through multi-stage or slow-thinking mechanisms. OneRec-Think~\cite{DBLP:journals/corr/abs-2510-11639} and R4ec~\cite{10.1145/3705328.3748068} introduce offline deliberation or dual-model collaboration schemes, such as Actor–Reflector architectures, to enhance reasoning capacity. In a different direction, RecZero~\cite{DBLP:journals/corr/abs-2510-23077} directly optimizes recommendation metrics via reinforcement learning (RL) methods such as GRPO or PPO, leading to the emergent formation of reasoning behaviors without explicit reasoning supervision.

Despite their effectiveness, these approaches largely rely on explicit reasoning representations, which introduce additional inference latency and computational overhead, and often depend on high-cost and noise-sensitive CoT supervision. This limitation motivates the exploration of more efficient reasoning mechanisms that do not require explicit textual reasoning signals.

\subsection{LLM Latent Reasoning}
Latent reasoning has been proposed as a promising alternative to explicit Chain-of-Thought (CoT) \cite{wei2022chain}, aiming to shift reasoning processes from the textual space to continuous latent representations. In the context of general-purpose LLMs, Deng et al.~\cite{DBLP:journals/corr/abs-2311-01460} compress horizontal reasoning steps into vertical model depth through distillation. COCONUT~\cite{DBLP:journals/corr/abs-2412-06769} performs implicit planning using continuous hidden states, while Huginn~\cite{DBLP:journals/corr/abs-2502-05171} scales test-time computation by unrolling recurrent blocks in latent space.
In recommender systems, early attempts have explored implicit computation as a means to capture user intent. LARES~\cite{DBLP:journals/corr/abs-2505-16865} and ReaRec~\cite{DBLP:journals/corr/abs-2503-22675} employ deep recurrent structures or reasoning position embeddings to model implicit preference evolution. Most closely related to our work, LatentR\textsuperscript{3}~\cite{zhang2025reinforced} introduces RL to generate continuous latent tokens, thereby removing the need for explicit reasoning supervision. 

However, its architecture relies on a single-head attention mechanism, effectively modeling user decisions through a single latent factor. Such a design may limit the model’s ability to capture multi-factor user preferences, such as trade-offs among price, brand, and functionality, motivating the need for more structured latent reasoning mechanisms.

\section{Problem Formulation}
Following prior work~\cite{you2025r,zhang2025reinforced}, we denote the collected recommendation dataset as $\mathcal{D}$, where each instance $(u, h, y) \in \mathcal{D}$ corresponds to a user–item interaction. Here, $u$ denotes a user, $h$ represents the user’s historical interaction sequence, and $y$ is the subsequent item with which the user interacts. Both the history $h$ and the target item $y$ are represented using textual metadata, such as item titles or descriptions.
To leverage LLMs, we reformulate the recommendation task as a natural language generation problem. For each instance $(u, h, y)$, we convert the historical sequence $h$  into a textual prompt $x$, and thus represent the data point as $(x, y)$, 
where $|x|$ denotes the token length of the prompt.

Since user preferences are implicitly embedded within the historical 
interactions, we do not directly ask the LLM to generate the next-item 
prediction. Instead, we introduce an intermediate reasoning process. 
Given a prompt $x$, the LLM first generates a latent reasoning output 
$r$, which we refer to as \emph{thoughts}, and subsequently produces the 
predicted item $\hat{y}$. Formally, this reasoning-augmented generation 
procedure can be expressed as:
\begin{equation}
    x 
    \xrightarrow{\text{LLM}} 
    r 
    \xrightarrow{\text{LLM}} 
    \hat{y},
\end{equation}

Our objective is to train the LLM to perform such 
recommendation-oriented reasoning directly from the dataset $\mathcal{D}$, 
without requiring any external supervision in the form of explicit 
chain-of-thought (CoT) annotations. In other words, the model must learn 
to generate effective intermediate reasoning $r$ in a self-supervised 
manner while optimizing the final prediction $\hat{y}$.

\section{Methodology}

\begin{figure*}[t]
    \centering
    {
        \includegraphics[width=\linewidth]{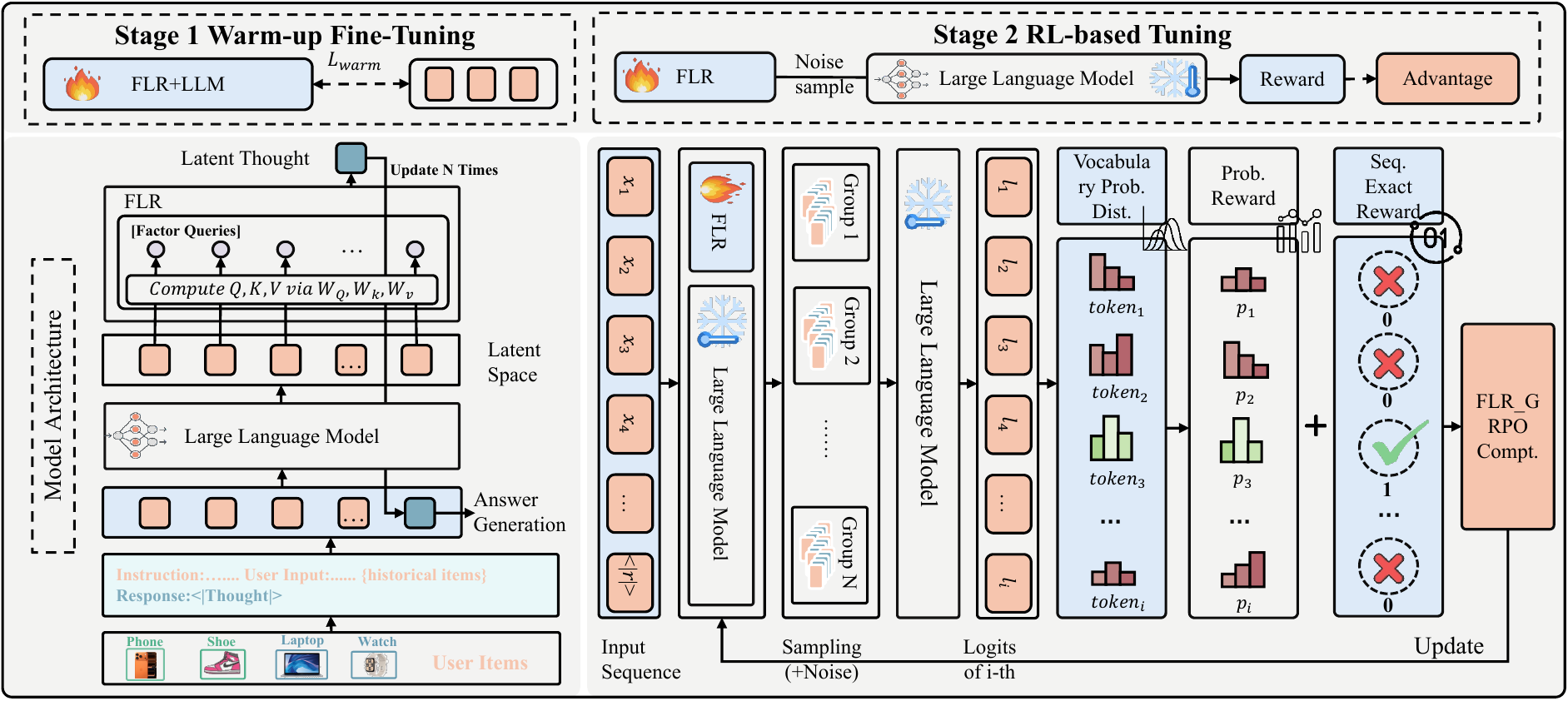}
    }
    \vspace{-2em}
    \caption{Illustration of the FLR architecture and its Two-Stage training paradigm.}
    \label{fig:latentflr}
\end{figure*}

The overall architecture of our proposed framework is illustrated in Figure \ref{fig:latentflr}. It depicts the end-to-end pipeline, starting from the iterative factorization of user intents in the latent space to the reinforcement learning-based optimization via Group Relative Policy Optimization (GRPO), which aligns the reasoning factors with the final recommendation objective.

\subsection{Factorized Latent Reasoning for Recommendation}\label{sec:flr_model}

\begin{figure}[t]
    \centering
    {
        \includegraphics[width=\linewidth]{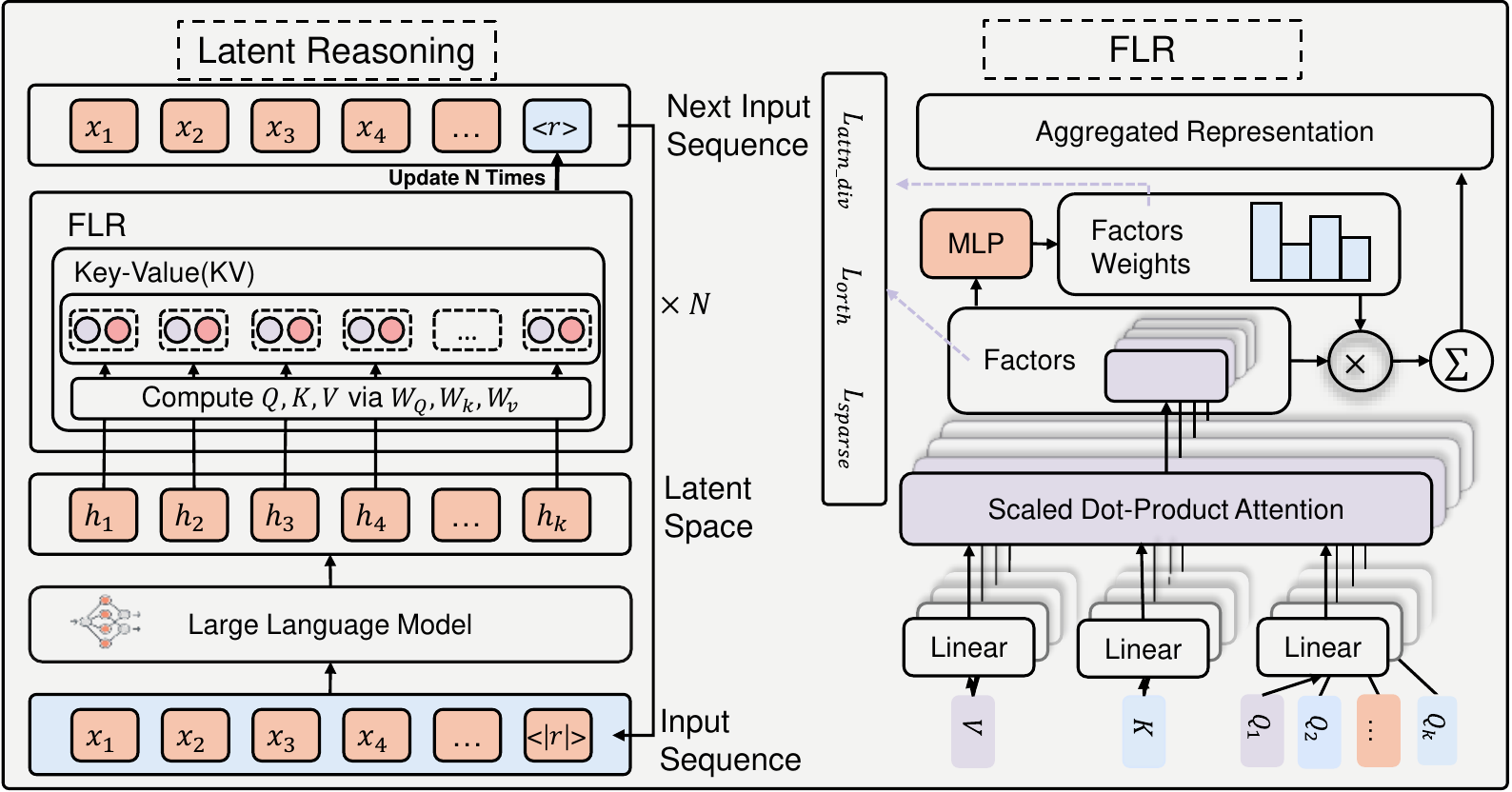}
    }
    \vspace{-2em}
    \caption{Illustration of the Factorized Latent Reasoning (FLR) module.} 
    \label{fig:flr}
\end{figure}

To capture the multifaceted and evolving preferences of users, we propose the Factorized Latent Reasoning (FLR) framework. FLR introduces an iterative refinement mechanism that disentangles user interests into distinct latent factors before generating the final recommendation. 

Figure \ref{fig:flr} provides a detailed visualization of this internal mechanism. As shown, the FLR module projects the context into multiple orthogonal spaces to capture diverse semantic facets, which are then dynamically weighted and aggregated to update the thought token.


\paragraph{Iterative Latent Refinement.}
Unlike standard sequential recommenders that process the interaction history in a single pass, FLR employs an iterative "thinking" process. We augment the user interaction sequence $\mathbf{x}$ by appending a special learnable token, \texttt{<|Thought|>}, to form the input $\tilde{\mathbf{x}} = [\mathbf{x}; \texttt{<|Thought|>}]$. Let $\mathbf{E}^{(0)}$ denote the initial embedding matrix. The reasoning process proceeds for $N$ iterations. In each step $n$(where $1 \le n \le N$), the LLM backbone $\Phi$ processes the current embeddings to produce hidden states $\mathbf{H}^{(n)} = \Phi(\mathbf{E}^{(n-1)})$. The aggregated reasoning vector $\mathbf{z}^{(n)}$, derived from the factorized attention module (detailed below), is used to perform an \textit{in-place update} of the thought token's embedding:
\begin{equation}
    \mathbf{E}^{(n)}[\text{pos}_{\texttt{<|Thought|>}}] \leftarrow \mathbf{z}^{(n)},
\end{equation}
This recursive update allows the model to progressively elaborate its understanding of the user's intent within a fixed sequence length.

\paragraph{Multi-Factor Disentanglement and Aggregation.}
The core of FLR is to extract $K$ distinct preference factors from the context $\mathbf{H}^{(n)}$. We maintain $K$ learnable query prototypes $\mathbf{Q}_f \in \mathbb{R}^{K \times D}$. To incorporate temporal dynamics, we apply Rotary Position Embeddings (RoPE) to the keys while projecting the hidden states. The factor-specific attention scores $\mathbf{A}^{(n)} \in \mathbb{R}^{K \times L_{\text{in}}}$ are computed as:
\begin{equation}
    \mathbf{A}^{(n)} = \text{softmax}\left( \frac{\mathbf{W}_q\mathbf{Q}_f (\text{RoPE}(\mathbf{W}_k\mathbf{H}^{(n)}))^\top}{\sqrt{D}} + \mathbf{M} \right),
\end{equation}
where $\mathbf{M}$ denotes the attention mask. The extracted factor representations $\mathbf{F}^{(n)} = \mathbf{A}^{(n)} \mathbf{V}^{(n)}$ are then dynamically aggregated. Recognizing that not all factors are equally relevant for every context, we employ an MLP-based gating mechanism to compute importance weights $\boldsymbol{\alpha}^{(n)}$:
\begin{equation}
    \mathbf{z}^{(n)} = \sum_{k=1}^{K} \alpha_k^{(n)} \mathbf{F}_k^{(n)}, \quad \text{with } \boldsymbol{\alpha}^{(n)} = \text{softmax}(\text{MLP}(\text{flatten}(\mathbf{F}^{(n)}))),
\end{equation}
The aggregated representation $\mathbf{z}^{(n)}$ is then used to update the embedding of the special thought token. After the final refinement step, the LLM performs next-item generation conditioned on the original prompt and the refined thought representation. 

\paragraph{Regularization Objectives.}
To ensure the FLR module learns a robust and interpretable reasoning structure, we introduce three regularization constraints. These objectives are designed to enforce global diversity in the latent space, distinct temporal receptive fields in the attention mechanism, and local specialization in the final decision-making process.

First, we minimize the \textit{Orthogonality Loss} to prevent \textbf{global degeneration}. Let $\hat{\mathbf{F}}_b$ denote the L2-normalized factor matrix where each row is scaled to unit length ($\|\hat{\mathbf{F}}_{b,k}\|_2 = 1$). The loss is defined as:
\begin{equation}
    \mathcal{L}_{\text{orth}} = \frac{1}{B} \sum_{b=1}^{B} \left\| \hat{\mathbf{F}}_b \hat{\mathbf{F}}_b^\top - \mathbf{I}_K \right\|_F^2,
\end{equation}
Mechanistically, this constraint forces latent vectors to span orthogonal subspaces, ensuring semantic disentanglement. For example, in a fashion recommendation scenario, it guarantees that while one factor ($F_1$) encodes \textit{"Visual Style"} (e.g., color, pattern), another factor ($F_2$) captures a distinct attribute like \textit{"Price Sensitivity"}. Without this global constraint, the model risks redundancy, where all factors merely replicate dominant signals (e.g., popularity) and fail to capture the multifaceted nature of user preferences.

Second, to prevent vision overlap and ensure factors capture dependencies across varying time scales, we apply an \textit{Attention Diversity Loss}:
\begin{equation}
    \mathcal{L}_{\text{div}} = \frac{2}{K(K-1)} \sum_{i < j} \text{cos}(\mathbf{A}_{i}, \mathbf{A}_{j}),
\end{equation}
Crucially, while orthogonality separates \textit{what} features are learned, this term constrains \textit{where} the model looks. By explicitly penalizing high similarity between attention maps, it differentiates the \textbf{temporal receptive fields} of factors. For instance, it guides one factor to focus on short-term triggers (e.g., a recent phone click implying a need for a case), while compelling another to attend to long-term periodic habits (e.g., monthly pet food purchases). This ensures the model maintains a comprehensive view of user history rather than myopically focusing solely on recent interactions.

Third, to encourage \textbf{local specialization} and decisive reasoning, we impose a \textit{Sparsity Loss} on the importance weights $\boldsymbol{\alpha}$:
\begin{equation}
    \mathcal{L}_{\text{sparse}} = -\frac{1}{B} \sum_{b=1}^{B} \sum_{k=1}^{K} \alpha_{b,k} \log(\alpha_{b,k} + \epsilon),
\end{equation}
Minimizing entropy pushes the weight distribution towards a one-hot vector, inducing a "winner-take-all" effect. This ensures that for any \textit{specific} prediction instance, the reasoning is driven by a single dominant factor. For example, if a user is exploring "Camping Gear", the decision should be dominated by the \textit{"Category Need"} factor, while irrelevant factors (like \textit{"Visual Style"} or accidental clicks) are effectively silenced. This prevents the final representation from becoming a noisy average of conflicting intents.

The final training objective integrates the recommendation loss $\mathcal{L}_{\text{rec}}$ with these structural constraints:
\begin{equation}
    \mathcal{L}_{\text{total}} = \mathcal{L}_{\text{rec}} + \lambda_1 \mathcal{L}_{\text{orth}} + \lambda_2 \mathcal{L}_{\text{div}} + \lambda_3 \mathcal{L}_{\text{sparse}},
\end{equation}
Balancing these objectives is non-trivial, as the optimal trade-off between global diversity and local specialization varies during training. Therefore, we employ an uncertainty-based weighting strategy to adaptively learn the coefficients $\lambda_1, \lambda_2, \lambda_3$. For the detailed formulation and initialization of this adaptive mechanism, please refer to Section 6.2 (Implementation Details).


\subsection{Learning Factorized Latent Reasoning via GRPO}\label{sec:grpo_finetuning}

Following the warm-up fine-tuning phase, we further align the Factorized Latent Reasoning (FLR) module with the recommendation objective using a GRPO-style reinforcement learning framework \cite{shao2024deepseekmath} inspired by LatentR3~\cite{zhang2025reinforced}. 
Different from LatentR3, which optimizes a monolithic latent reasoning representation, our approach performs latent-space exploration and policy optimization over factorized reasoning representations. 
We further adapt the reward design to the generative recommendation setting by combining token-level confidence with sequence-level exact-match feedback. 
During this phase, we freeze the LLM backbone for parameter efficiency and update only the FLR parameters $\theta$.

\paragraph{Latent Space Exploration Strategy.}
Standard reinforcement learning in language modeling typically relies on autoregressive sampling to generate diverse completion trajectories, which incurs high computational costs. To mitigate this, we introduce a computationally efficient \textit{latent perturbation} mechanism. Instead of sampling tokens, we inject Gaussian noise directly into the \texttt{<|Thought|>} token embedding to induce diverse reasoning paths. Formally, given an input sequence $x$, the reasoning embedding $\mathbf{e}_{\texttt{<|T|>}}$ for the $i$-th sample in a group of size $G$ is perturbed as:
\begin{equation}
    \tilde{\mathbf{e}}^{(i)}_{\texttt{<|T|>}} = \mathbf{e}_{\texttt{<|T|>}} + \boldsymbol{\epsilon}_i, \quad \text{where } \boldsymbol{\epsilon}_i \sim \begin{cases} \mathbf{0}, & \text{if } i=1 \text{ (Baseline)} \\ \mathcal{N}(\mathbf{0}, \sigma^2 \mathbf{I}), & \text{otherwise} \end{cases},
\end{equation}
Crucially, the first sample ($i=1$) remains unperturbed ($\boldsymbol{\epsilon}_1 = \mathbf{0}$), serving as a stable baseline for variance reduction during advantage estimation.

\paragraph{Hybrid Reward Formulation.}
To address the issue of reward sparsity in sequential recommendation, we design a hybrid reward function that combines dense probabilistic feedback with discrete success signals. The reward $r$ for a generated response $y$ is defined as:
\begin{equation}
    r(x, y) = \alpha \cdot \underbrace{\frac{1}{L} \sum_{t=1}^{L} \log \pi_\theta(y_t | x, y_{<t})}_{\text{Token Confidence}} + \beta_r \cdot \underbrace{\mathbb{I}(\hat{y} = y)}_{\text{Exact Match}},
\end{equation}
where $\mathbb{I}(\cdot)$ is the indicator function for exact sequence matching, and $\alpha, \beta_r$ are hyperparameters balancing the two terms. This formulation encourages the model to generate high-confidence predictions while ultimately optimizing for the correct item recommendation.

\paragraph{Group-Relative Advantage Estimation.}
We estimate the advantage function using a group-based baseline subtraction method. For the $i$-th sample within a group, the advantage $\hat{A}_i$ is computed relative to the noise-free baseline reward $r_{\text{base}}$ (from sample $i=1$):
\begin{equation}
    \hat{A}_i = \frac{r_i - r_{\text{base}}}{\|\mathbf{r}_{2:G} - r_{\text{base}}\|_2 + \epsilon},
\end{equation}
Here, we utilize L2 normalization over the group deviations to stabilize gradient magnitudes, which we empirically found to be more robust than standard z-score normalization for our hybrid reward scale.

\paragraph{Optimization Objective.}
The policy is optimized using a clipped surrogate objective constrained by a reverse KL divergence penalty. Let $\rho_t(\theta) = \pi_\theta(y_t|x) / \pi_{\theta_{\text{old}}}(y_t|x)$ denote the probability ratio. The per-token objective $\mathcal{L}_{\text{token}}$ is formulated as:
\begin{equation}
    \mathcal{L}_{\text{token}} = \min\left( \rho_t \hat{A}_t, \text{clip}(\rho_t, 1-\epsilon_l, 1+\epsilon_h) \hat{A}_t \right) - \beta_{\text{KL}} D_{\text{KL}},
\end{equation}
where we use asymmetric clipping bounds ($\epsilon_l=0.2, \epsilon_h=0.28$) to encourage stable improvements. The KL penalty is approximated via the reverse form $D_{\text{KL}} \approx e^{\Delta} - \Delta - 1$ (with $\Delta = \log \pi_{\text{ref}} - \log \pi_\theta$) to prevent the policy from deviating excessively from the reference model $\pi_{\text{ref}}$. 

Finally, to maintain the interpretability of the learned factors during fine-tuning, we integrate the FLR regularization terms, yielding the total objective:
\begin{equation}
    \mathcal{L}_{\text{total}} = \mathbb{E}_{\mathcal{D}} \left[ -\mathcal{L}_{\text{token}} \right] + \lambda_1 \mathcal{L}_{\text{orth}} + \lambda_2 \mathcal{L}_{\text{div}} + \lambda_3 \mathcal{L}_{\text{sparse}}.
\end{equation}

\section{Experiments}

\subsection{Research Questions}
In this section, we aim to answer the following research questions (\textbf{RQs}):
\begin{itemize}[leftmargin=*]
    \item \textbf{RQ1:} How does the performance of FLR compare with other baselines across the datasets in different experiment settings?
    \item \textbf{RQ2:} What is the impact of the key design components in FLR on overall recommendation performance?
    \item \textbf{RQ3:} How does the proposed GRPO mechanism impact the performance of FLR, and does it offer superior robustness compared to a generic reinforcement learning strategy?
    \item \textbf{RQ4:} How does the model behave regarding internal attention mechanisms?
    \item \textbf{RQ5:} How does the model perform on difficult (unpopular) items?
    \item \textbf{RQ6:} How sensitive is FLR to the number of latent reasoning factors, and what factor configuration achieves the best recommendation performance?
    \item \textbf{RQ7:} Is the inference overhead acceptable for practical use?
\end{itemize}

\subsection{Experiment Setup}


\textbf{Datasets.} We conduct our experiments on four domain-specific subsets from the Amazon review dataset~\cite{ni2019justifying}, namely \textit{Toys}, \textit{CDs}, \textit{Games}, and \textit{Instruments}. 
To ensure high-quality data suitable for LLM training, we employ a dynamic temporal sliding window strategy for data selection.  Specifically, we initialize the data window from October 2017 to October 2018 and apply 5-core filtering. 
To guarantee sufficient data density, if the number of unique items remains below 5,000, we iteratively extend the start date backwards by 3-month intervals (e.g., to July 2017) while keeping the end date fixed, until the threshold is met.

Following this selection process, we partition the dataset into training, validation, and testing sets according to a chronological 8:1:1 split. 
Furthermore, to maintain consistency with baseline models and analyze temporal patterns effectively, the maximum sequence length is standardized to 10. 
Table~\ref{tab:statistics} summarizes the detailed statistics of the processed datasets.

\begin{table}[h]
    \centering
    \vspace{-1em}
    \caption{Dataset statistics.}
    \vspace{-1em}
    \label{tab:statistics}
    \setlength{\tabcolsep}{8pt} 
    \begin{tabular}{lcccc}
        \toprule
        \textbf{Dataset} & \textbf{Train} & \textbf{Valid} & \textbf{Test} & \textbf{Item} \\
        \midrule
        Toys        & 53,898 & 6,737 & 6,738 & 6,299 \\
        CDs         & 49,251 & 6,156 & 6,158 & 5,841 \\
        Games       & 75,175 & 9,397 & 9,397 & 5,308 \\
        Instruments & 66,500 & 8,312 & 8,313 & 5,030 \\
        \bottomrule
    \end{tabular}
    \vspace{-1em}
\end{table}

\textbf{Evaluation Metrics.}
We employ two standard top-$K$ recommendation metrics, \textit{Normalized Discounted Cumulative Gain@K} (NDCG@K) and \textit{Hit Ratio@K} (HR@K), to evaluate performance.
We report results for $K \in \{5, 10\}$.
For brevity, HR@5 and NDCG@5 are denoted as H@5 and N@5, respectively.

\textit{Ranking Strategy.}
For traditional retrieval-based baselines, item rankings are produced following their original scoring protocols.
For LLM-based generative recommenders, following \cite{zhang2025reinforced}, we adopt a constrained generative ranking protocol.
Specifically, we construct a prefix trie from the textual titles of all candidate items in the catalog and constrain beam search such that every generated sequence corresponds to a valid catalog item.
In FLR, during the reasoning phase, the \texttt{<|Thought|>} token is iteratively updated by aggregating latent preference factors.
During decoding, the LLM generates valid item titles under the catalog constraint, and the beam-search sequence scores are used to form the top-$K$ recommendation list.
Since the prefix trie is constructed from all items in the dataset, the decoding space covers the full candidate set while avoiding post-hoc matching between free-form generations and catalog items.
Following \cite{zhang2025reinforced}, statistical significance is verified using a paired t-test with $p<0.05$.

\textbf{Implementation Details.} Following the setting in \cite{zhang2025reinforced}, we employ Qwen2.5-1.5B \cite{ahmed2025qwen} as the backbone. For Supervised Fine-Tuning (SFT), we utilize the AdamW optimizer \cite{loshchilov2017decoupled}, with the learning rate tuned within $\{3\text{e-}4, 5\text{e-}5, 1\text{e-}5\}$.
For the latent reasoning module, the number of factors $K$ is selected via grid search from $\{2, 3, 4\}$; based on validation, we set $K=3$ for \textit{CDs} and $K=4$ for \textit{Toys}, \textit{Games}, and \textit{Instruments}. 
The number of reasoning iterations $T$ is set to $2$, consistent with previous work \cite{zhang2025reinforced}.
To balance the multi-objective optimization, we adopt an uncertainty-based weighting strategy \cite{kendall2018multi} to adaptively learn $\lambda_1, \lambda_2, \lambda_3$, where each weight is computed as $\lambda = \frac{1}{2 \exp(s)}$ ($s$ initialized to zero).
For the post-training phase, the learning rate is searched within $\{1\text{e-}4, 5\text{e-}4, 1\text{e-}5, 5\text{e-}5\}$. Reward hyperparameters are tuned within $[0.1, 1.0]$, with optimal values set to $\alpha=0.1$ and $\beta=1.0$.
All experiments were conducted on a cluster with 2 NVIDIA A100 GPUs. To ensure statistical robustness, all reported results are averaged over five independent runs with different random seeds.

\textbf{Baselines.} 
We select various representative and/or state-of-the-art recommendation models as baselines. (1) Traditional sequential models, including \textbf{Caser} \cite{DBLP:conf/wsdm/TangW18} models the sequence of recent user interactions as an image in time and latent spaces, utilizing convolutional filters to capture high-order sequential patterns and general preferences. \textbf{GRU4Rec} \cite{DBLP:conf/cikm/HidasiK18}: an improved RNN-based framework that optimizes session-based recommendation via gradient-stable ranking losses and additional negative sampling to maximize Recall and MRR.
\textbf{SASRec} \cite{DBLP:conf/icdm/KangM18}: employs a self-attention mechanism to capture long-term sequential semantics, effectively balancing the parsimony of Markov Chains and the expressiveness of RNNs to achieve superior performance and efficiency across datasets of varying sparsity. (2) LLM-based approaches, including 
\textbf{Base}, directly utilize the generative and reasoning capabilities of LLMs to predict the next item by processing the user's historical interaction sequence as textual input. \textbf{CoT} \cite{tsai-etal-2024-leveraging} investigates the application of Chain-of-Thought reasoning in subjective personalized recommendations and introduces an automated framework to evaluate the coherence and faithfulness of reasoning paths without human supervision. \textbf{AlphaRec} \cite{DBLP:conf/iclr/Sheng0ZCWC25}, this work reveals the homomorphism between language and recommendation spaces, demonstrating that a collaborative filtering model constructed purely from LM-encoded item text without ID embeddings can outperform leading ID-based baselines. \textbf{BIGRec} \cite{DBLP:journals/tors/BaoZWZYLCFT25} introduces a bi-step grounding paradigm that first fine-tunes LLMs to generate meaningful intermediate tokens and subsequently maps them to actual items, enabling effective ranking capabilities over the full candidate set. \textbf{LatentR$^3$} \cite{zhang2025reinforced} proposes a reinforced latent reasoning framework that optimizes compact latent tokens via a two-stage training strategy combining supervised initialization with a modified GRPO-based reinforcement learning, thereby achieving efficient, high-performance recommendation without explicit chain-of-thought supervision.

\subsection{Main Results (RQ1)}

As shown in Table \ref{tab:main_results}, FLR achieves state-of-the-art performance across all datasets, validating the effectiveness of the constrained multi-head attention mechanism. Specifically, FLR significantly outperforms traditional models, achieving a maximum relative improvement of 84.6\%, and exceeds untuned LLMs by over 265.4\%, confirming the synergistic effect of combining LLM semantics with structured latent factors. Meanwhile, it maintains a substantial lead over LLM-based baselines such as AlphaRec, exemplified by a 16.4\% gain in N@5 on the Games dataset, and consistently surpasses \textit{LatentR\textsuperscript{3}} across all metrics on Toys, CDs, and Games, demonstrating its superior capability in capturing diverse user intents. Furthermore, the performance gains are closely related to domain characteristics. FLR brings larger improvements on interest-driven domains such as Toys, CDs, and Games, where user preferences are often diverse and multi-faceted, highlighting the benefit of disentangling latent reasoning into multiple preference factors.
Conversely, the narrower margins on the Instruments dataset suggest that, due to its professional nature and explicit decision logic, the underlying intent space is lower-dimensional, where single-head reasoning suffices. This highlights that the granularity of latent reasoning must align with the decision complexity of the recommendation domain.

\begin{table*}[t]
\centering
\caption{Top-$N$ recommendation performance of FLR versus baselines. "RI" indicates the
relative improvement of FLR over each baseline across all datasets and metrics.
Best results are in \textbf{bold}, second-best are \underline{underlined}. \textit{Improv} denotes the relative improvement of FLR over the strongest baseline. All results are averaged over five runs with different seeds, reported as mean $\pm$ standard deviation. 
We conduct paired t-tests over five independent runs. Statistically significant improvements over the strongest baseline are marked with * at $p < 0.05$.
}
\label{tab:main_results}
\vspace{-1em}
\small  

\setlength{\tabcolsep}{5pt}
\begin{tabular}{l|l|ccc|ccccccc}
\toprule
\multirow{2}{*}{\textbf{Dataset}} & \multirow{2}{*}{\textbf{Methods}} & \multicolumn{3}{c|}{\textbf{Traditional}} & \multicolumn{6}{c}{\textbf{LLM-based}} & {\multirow{2}{*}{\textbf{Improv (\%)}}}\\
\cmidrule(lr){3-5} \cmidrule(lr){6-11}
& \textbf{Metrics} & \textbf{Caser} & \textbf{GRU4Rec} & \textbf{SASRec} & \textbf{Base} & \textbf{COT} & \textbf{AlphaRec} & \textbf{BIGRec} & \textbf{LatentR$^3$} & \textbf{FLR} \\
\midrule
\midrule
\multirow{4}{*}{Toys} 
& H@5 & 0.0251 & 0.0417 & 0.0601 & 0.0203 & 0.0261 & 0.0579 & 0.0701 & \underline{0.0781} & \textbf{0.0814$\pm$0.0006}$^*$ & 4.23\% \\
& H@10 & 0.0384 & 0.0564 & 0.0760 & 0.0359 & 0.0496 & 0.0893 & 0.0931 & \underline{0.1068} & \textbf{0.1077$\pm$0.0003} & 0.84\% \\
& N@5 & 0.0170 & 0.0305 & 0.0458 & 0.0128 & 0.0153 & 0.0347 & 0.0508 & \underline{0.0579} & \textbf{0.0611$\pm$0.0008}$^*$ & 5.53\% \\
& N@10 & 0.0214 & 0.0352 & 0.0510 & 0.0178 & 0.0229 & 0.0448 & 0.0582 & \underline{0.0674} & \textbf{0.0695$\pm$0.0004}$^*$ & 3.12\% \\
\midrule
\multirow{4}{*}{CDs} 
& H@5 & 0.0469 & 0.0481 & 0.0841 & 0.0195 & 0.0302 & 0.0479 & 0.0757 & \underline{0.0816} & \textbf{0.0857$\pm$0.0008}$^*$ & 5.02\% \\
& H@10 & 0.0689 & 0.0669 & \textbf{0.1054} & 0.0252 & 0.0406 & 0.0774 & 0.0929 & 0.0992 & \underline{0.1013$\pm$0.0007}$^*$ & 2.12\% \\
& N@5 & 0.0312 & 0.0365 & 0.0622 & 0.0148 & 0.0213 & 0.0278 & 0.0616 & \underline{0.0662} & \textbf{0.0689$\pm$0.0006}$^*$ & 4.08\% \\
& N@10 & 0.0382 & 0.0425 & 0.0691 & 0.0167 & 0.0246 & 0.0373 & 0.0672 & \underline{0.0719} & \textbf{0.0741$\pm$0.0002}$^*$ & 3.06\% \\
\midrule
\multirow{4}{*}{Games} 
& H@5 & 0.0324 & 0.0322 & 0.0416 & 0.0236 & 0.0120 & 0.0558 & 0.0461 & \underline{0.0593} & \textbf{0.0639$\pm$0.0012}$^*$ & 7.76\% \\
& H@10 & 0.0538 & 0.0517 & 0.0633 & 0.0311 & 0.0194 & 0.0893 & 0.0709 & \underline{0.0889} & \textbf{0.0908$\pm$0.0006}$^*$ & 2.14\% \\
& N@5 & 0.0211 & 0.0207 & 0.0280 & 0.0190 & 0.0082 & 0.0397 & 0.0334 & \underline{0.0419} & \textbf{0.0462$\pm$0.0005}$^*$ & 10.26\% \\
& N@10 & 0.0280 & 0.0270 & 0.0350 & 0.0214 & 0.0105 & 0.0515 & 0.0414 & \underline{0.0515} & \textbf{0.0548$\pm$0.0003}$^*$ & 6.41\% \\
\midrule
\multirow{4}{*}{Instruments} 
& H@5 & 0.0781 & 0.0766 & 0.0793 & 0.0296 & 0.0261 & 0.0813 & 0.0938 & \underline{0.1029} & \textbf{0.1032$\pm$0.0002}$^*$ & 0.29\% \\
& H@10 & 0.0977 & 0.0960 & 0.0950 & 0.0411 & 0.0452 & 0.1051 & 0.1158 & \underline{0.1214} & \textbf{0.1248$\pm$0.0005}$^*$ & 2.80\% \\
& N@5 & 0.0564 & 0.0630 & 0.0708 & 0.0154 & 0.0135 & 0.0564 & 0.0807 & \underline{0.0882} & \textbf{0.0886$\pm$0.0004}$^*$ & 0.45\% \\
& N@10 & 0.0627 & 0.0692 & 0.0758 & 0.0192 & 0.0199 & 0.0640 & 0.0879 & \underline{0.0941} & \textbf{0.0955$\pm$0.0008}$^*$ & 1.49\% \\
\midrule
\multirow{1}{*}{ }
& RI & 84.6{\%} & 66.9{\%} & 32.6{\%} & 265.4{\%} & 244.5{\%} & 38.2{\%} & 16.5{\%} & 3.2{\%} & - & - \\
\bottomrule
\end{tabular}
\end{table*}

\subsection{Ablation Study (RQ2)}

\begin{figure}[t]
    \centering
    {
        \includegraphics[width=0.95\linewidth]{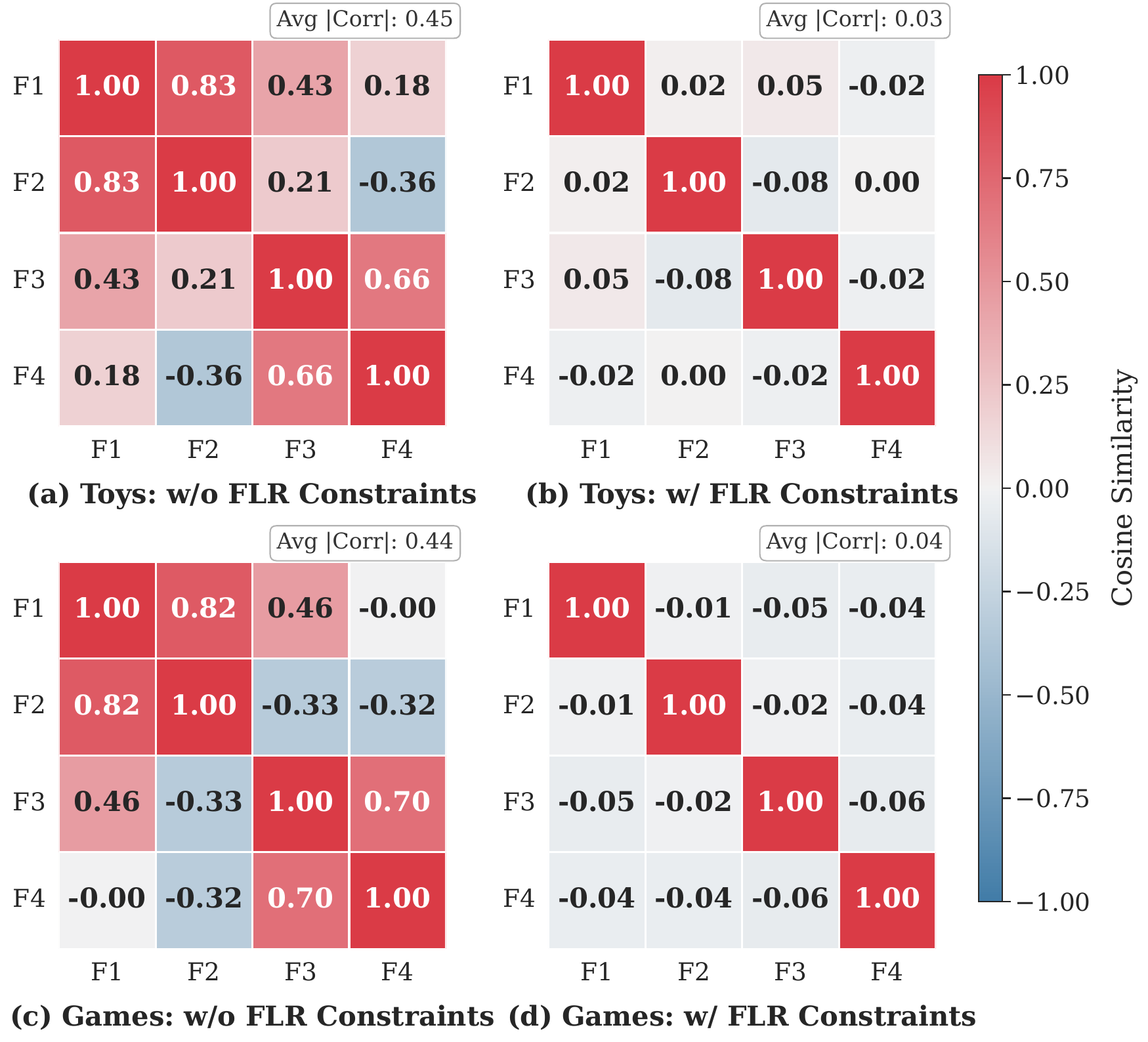}
    }
    \vspace{-1em}
    \caption{Visualization of factor disentanglement on Toys and Games datasets. (a, c) \textbf{Unconstrained Setting}: Latent factors exhibit high redundancy (Avg |Corr| $\approx$ 0.44), indicating mode collapse. (b, d) \textbf{With FLR Constraints}: Factors become highly orthogonal and independent (Avg |Corr| $\approx$ 0.03), validating the efficacy of the proposed regularization.} 
    \label{fig:heatmap}
    \vspace{-1em}
\end{figure}

To assess the efficacy of the proposed Factorized Latent Reasoning (FLR) module and its regularization components, we conduct a systematic ablation study on the Games and Toys datasets. We isolate the contribution of three loss functions: Orthogonality ($\mathcal{L}_{orth}$), Attention Diversity ($\mathcal{L}_{attn\_div}$), and Sparsity ($\mathcal{L}_{sparse}$). Quantitative results are detailed in Table \ref{tab:ablation}, while the qualitative analysis of factor disentanglement is visualized in Figure \ref{fig:heatmap}.




\subsubsection{Quantitative Analysis}

Table \ref{tab:ablation} confirms that the combined regularization strategy significantly outperforms individual constraints. The unconstrained baseline consistently yields the poorest results across all metrics, exemplified by a 7.2\% lag behind the full model in H@5 on the Games dataset. This provides empirical evidence that simply scaling attention heads without structural constraints leads to mode collapse and feature redundancy. Single constraints prove insufficient. The sparse-only variant lags notably on the Toys dataset; specifically, the full model achieves a 3.55\% improvement in N@5 over the sparse-only counterpart, suggesting that enforcing sparsity on entangled factors suppresses valid signals. Similarly, while the orth-only variant improves semantic distinctness, it remains constrained by temporal "attention overlap", failing to capture long-term signals compared to the full model on the Games dataset. The full model achieves the best performance on 7 out of 8 metrics, validating the synergistic effect among the objectives: $\mathcal{L}_{orth}$ and $\mathcal{L}_{div}$ construct a diverse factor pool, enabling $\mathcal{L}_{sparse}$ to effectively filter noise. Notably, on the Toys dataset, the orth+sparse variant performs exceptionally well on H@10, which is attributed to the domain's static category preferences. Nevertheless, the full model retains superiority in the stricter ranking metric N@5 of \textbf{0.0611}, demonstrating superior general robustness.

\subsubsection{Qualitative Analysis of Factor Disentanglement}

To empirically verify the "Global Non-Degeneration" property of FLR, we visualize the cosine similarity matrices of the learned latent factors in Figure \ref{fig:heatmap}. Under the unconstrained setting shown in Figures \ref{fig:heatmap}(a) and \ref{fig:heatmap}(c), the model exhibits significant off-diagonal heat, with an average absolute correlation of approximately \textbf{0.44}. This visually corroborates the phenomenon of global degeneration: in the absence of regularization, multiple reasoning heads tend to converge into similar semantic subspaces. Mechanistically, this implies that the model is wasting its capacity by learning redundant features, thereby failing to provide diverse viewpoints for recommendations. In stark contrast, the full FLR model (Figures \ref{fig:heatmap}(b) and \ref{fig:heatmap}(d)) displays a strictly diagonal structure, where the average correlation drops to \textbf{0.03}. This confirms that our regularization strategy successfully forces the latent space to span orthogonal dimensions. Crucially, this disentanglement serves as the prerequisite for the synergy observed in the quantitative results (Table \ref{tab:ablation}). By ensuring via $\mathcal{L}_{orth}$ that each factor operates as an independent specialist, we enable the sparsity constraint ($\mathcal{L}_{sparse}$) to effectively identify and select the single most relevant factor for each prediction, rather than choosing from a noisy mixture of entangled signals.

\begin{table*}[t]
\centering
\vspace{-1em}
\caption{Ablation study results on Games and Toys datasets. "None" represents the FLR model without any regularization constraints ($\mathcal{L}_{orth}, \mathcal{L}_{div}, \mathcal{L}_{sparse}$). Bold values denote the best results.}
\label{tab:ablation}
\vspace{-1em}
\small
\setlength{\tabcolsep}{9pt}
\begin{tabular}{l|cccc|cccc}
\toprule
\multirow{2}{*}{\textbf{Methods}} & \multicolumn{4}{c|}{\textbf{Games}} & \multicolumn{4}{c}{\textbf{Toys}} \\
\cmidrule(lr){2-5} \cmidrule(lr){6-9}
& \textbf{H@5} & \textbf{H@10} & \textbf{N@5} & \textbf{N@10} & \textbf{H@5} & \textbf{H@10} & \textbf{N@5} & \textbf{N@10} \\
\midrule
None & 0.0596 & 0.0882 & 0.0435 & 0.0528 & 0.0779 & 0.1065 & 0.0585 & 0.0678 \\
\midrule
attn\_div & 0.0627 & 0.0886 & 0.0450 & 0.0533 & 0.0822 & 0.1073 & 0.0600 & 0.0681 \\
orth & 0.0625 & 0.0893 & 0.0460 & 0.0546 & 0.0793 & 0.1089 & 0.0594 & 0.0690 \\
sparse & 0.0631 & 0.0895 & 0.0458 & 0.0542 & 0.0782 & 0.1072 & 0.0590 & 0.0683 \\
attn\_div+orth & 0.0623 & 0.0898 & 0.0456 & 0.0545 & 0.0801 & 0.1069 & 0.0593 & 0.0678 \\
attn\_div+sparse & 0.0625 & 0.0902 & 0.0455 & 0.0544 & 0.0815 & 0.1076 & 0.0607 & 0.0691 \\
orth+sparse & 0.0607 & 0.0879 & 0.0443 & 0.0531 & 0.0794 & \textbf{0.1100} & 0.0586 & 0.0685 \\
attn\_div+orth+sparse & \textbf{0.0639} & \textbf{0.0908} & \textbf{0.0462} & \textbf{0.0548} & \textbf{0.0814} & 0.1077 & \textbf{0.0611} & \textbf{0.0695} \\
\bottomrule
\end{tabular}
\end{table*}

\subsection{Ablation Study on GRPO Mechanism (RQ3)}
\setlength{\textfloatsep}{9pt}
\begin{table}[t]
\small
\centering
\caption{Impact of our GRPO Strategies. Specifically, LR-GRPO is the improved GRPO method from LatentR$^3$, whereas FLR-GRPO is our proposed RL approach.}
\label{tab:AblationGRPO}
\vspace{-1em}
\small  
\setlength{\tabcolsep}{4pt}  
\begin{tabular}{ll|c|cccccccc}
\toprule
\textbf{Dataset} & \textbf{Metrics} & \textbf{LatentR$^3$} & \textbf{FLR} & \textbf{LR-GRPO} & \textbf{FLR-GRPO}\\
\midrule
\multirow{4}{*}{Toys} 
& H@5 & 0.0781  & 0.0814 & \underline{0.0815} & \textbf{0.0823} \\
& H@10 & 0.1068  & 0.1077 & \underline{0.1080} & \textbf{0.1125} \\
& N@5 & 0.0579 & \underline{0.0611} & 0.0601 & \textbf{0.0621} \\
& N@10 & 0.0674 & \underline{0.0695} & 0.0693 & \textbf{0.0722} \\
\midrule
\multirow{4}{*}{CDs} 
& H@5 & 0.0816 & 0.0857 & \underline{0.0859} & \textbf{0.0871} \\
& H@10 & 0.0992 & 0.1013 & \underline{0.1015} & \textbf{0.1024} \\
& N@5 & 0.0662 & 0.0689 & \underline{0.0689} & \textbf{0.0698} \\
& N@10 & 0.0719 & \underline{0.0741} & 0.0740 & \textbf{0.0751} \\
\midrule
\multirow{4}{*}{Games} 
& H@5 & 0.0593 & \underline{0.0639} & 0.0621 & \textbf{0.0645} \\
& H@10 & 0.0889 & \underline{0.0908} & 0.0906 & \textbf{0.0926} \\
& N@5 & 0.0419 & \underline{0.0462} & 0.0454 & \textbf{0.0468} \\
& N@10 & 0.0515 & \underline{0.0548} & 0.0546 & \textbf{0.0552} \\
\midrule
\multirow{4}{*}{Instruments} 
& H@5 & 0.1029 & \underline{0.1032} & 0.1028 & \textbf{0.1038} \\
& H@10 & 0.1214 & \underline{0.1248} & 0.1246 & \textbf{0.1254} \\
& N@5 & 0.0882 & \underline{0.0886} & 0.0885 & \textbf{0.0893}\\
& N@10 & 0.0941 & 0.0955 & \underline{0.0955} & \textbf{0.0964}\\
\midrule
\end{tabular}
\end{table}

\begin{figure}[t]
    \centering
    {
        \includegraphics[width=0.95\linewidth]{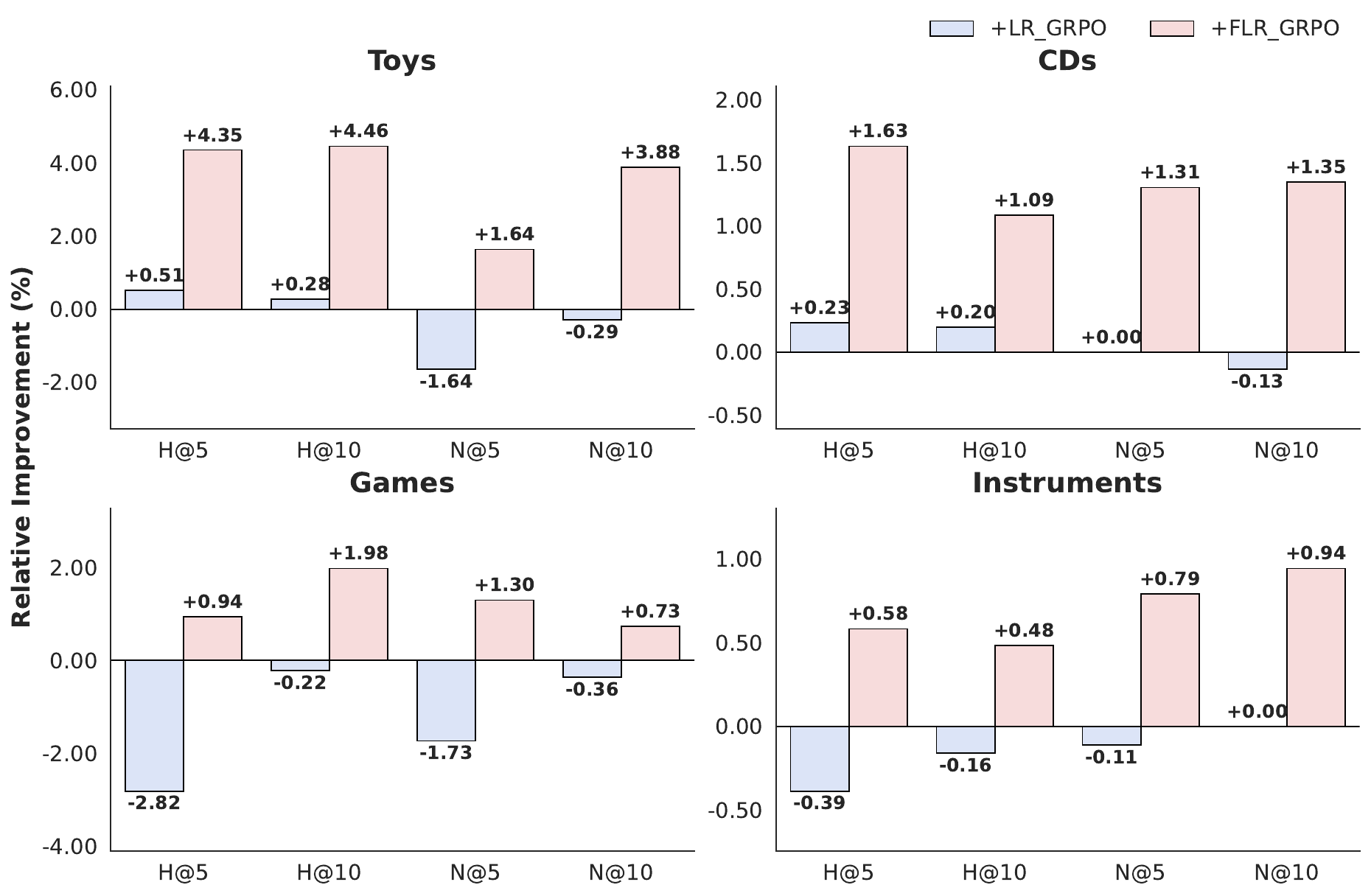}
    }
    \vspace{-1em}
    \caption{Relative performance improvement of LR-GRPO and FLR-GRPO over the FLR baseline. The comparison is conducted across four datasets (Toys, CDs, Games, Instruments) in terms of HR@K and NDCG@K. The baseline (0\%) represents the performance of the FLR model.}
    \label{fig:grpoimpro}
    \vspace{-1em}
\end{figure}

To rigorously evaluate the contribution of the proposed GRPO mechanism, we conduct a fine-grained comparison across four datasets. Figure~\ref{fig:grpoimpro} visualizes the relative performance improvement of the generic reinforcement learning variant (\texttt{+LR-GRPO}) and our proposed method (\texttt{+FLR-GRPO}) against the FLR baseline. The results reveal distinct performance patterns:

\textbf{(i) Instability of Generic GRPO (Negative Transfer).} 
As observed in the bottom row of Figure~\ref{fig:grpoimpro}, simply applying generic GRPO introduces optimization instability. On the \textit{Games} dataset, \texttt{+LR-GRPO} (gray bars) exhibits varying degrees of performance degradation across all metrics. Notably, it leads to a -2.82\% drop in H@5 and a -1.73\% drop in N@5 compared to the baseline. Similarly, on the \textit{Instruments} dataset, \texttt{+LR-GRPO} fails to provide positive gains, resulting in a slight decline (e.g., -0.39\% in H@5).
This phenomenon suggests that without a disentangled reasoning structure, the coarse-grained reward signal may act as noise, causing the model to diverge from the optimal policy.

\textbf{(2) Robustness and Superiority of FLR-GRPO.} In contrast, our proposed \texttt{+FLR-GRPO} (red bars) demonstrates superior robustness, consistently achieving positive improvements across all 16 metric-dataset combinations. \textbf{Reversing Negative Transfer:} On the difficult \textit{Games} dataset, our method successfully reverses the negative trend of the generic variant, turning the -2.82\% loss into a +0.94\% gain in H@5. \textbf{Significant Gains:} On the \textit{Toys} dataset (top-left), our method achieves the most substantial improvements. Specifically, it boosts H@10 by +4.46\% and N@10 by +3.88\%, significantly outperforming the generic variant.
These results strongly validate that the effectiveness of GRPO relies on the synergistic design with the FLR backbone. By aligning the reward mechanism with specific latent reasoning heads, \texttt{+FLR-GRPO} ensures stability in difficult domains (\textit{Games}) while maximizing potential in consistent domains (\textit{Toys}).


\begin{figure*}[t]
    \centering
    {
        \includegraphics[width=0.95\linewidth]{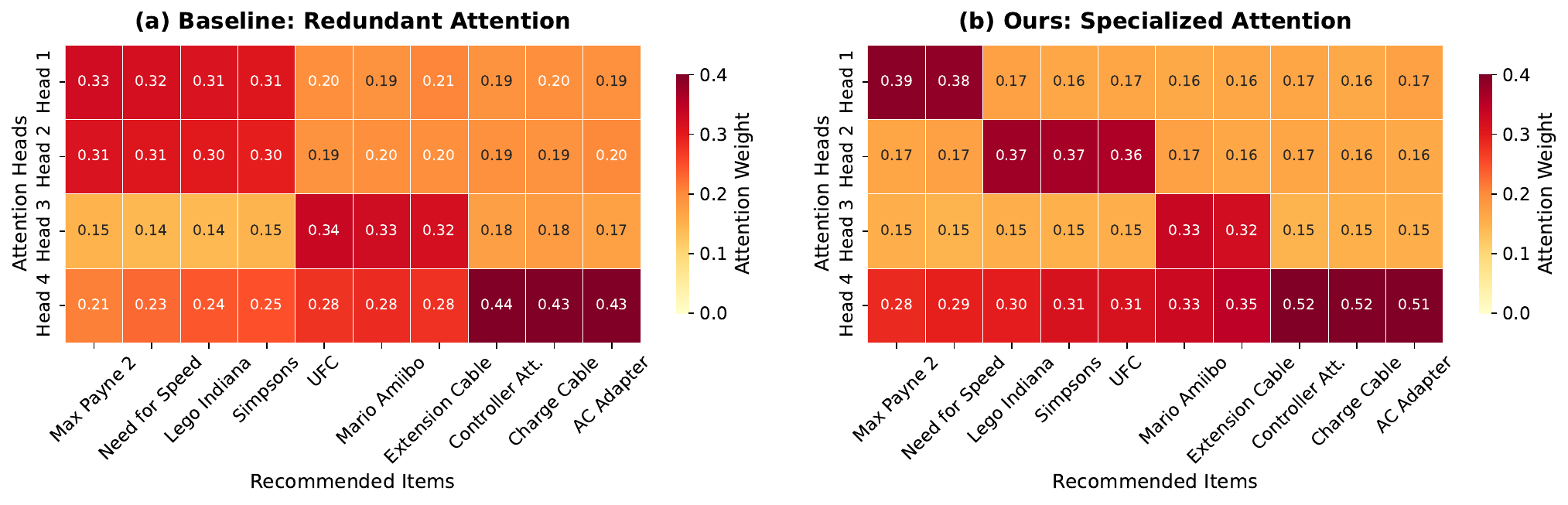}
    }
    \vspace{-1em}
    \caption{Visualization of attention patterns on Amazon Video Games. The x-axis shows recommended items, and the y-axis shows latent attention heads. (a) The pre-trained baseline exhibits redundant attention. (b) FLR-GRPO demonstrates category-aware specialization.}
    \label{fig:ItemAttentionMap}
\end{figure*}

\subsection{Attention Disentanglement (RQ4)}

To validate how FLR-GRPO enhances interpretability, we visualize the multi-head attention weights for a representative user from the Video Games dataset (Figure \ref{fig:ItemAttentionMap}).

\textbf{Analysis of Baseline Redundancy.}
As shown in Figure \ref{fig:ItemAttentionMap}(a), the pre-trained baseline exhibits significant \textit{attention redundancy}. Heads 1 and 2 display highly correlated activation patterns, both heavily prioritizing dominant categories (Action/Racing games like \textit{Max Payne 2}) while neglecting niche needs. This "mode collapse" indicates that multiple heads are effectively collapsing into a single semantic channel.

\textbf{Emergence of Specialized Heads.}
In contrast, Figure \ref{fig:ItemAttentionMap}(b) reveals that FLR-GRPO induces a spontaneous specialization of heads into distinct semantic roles. We observe a clear functional disentanglement:
\textbf{Head 1} maintains a focus on \textit{Core Gameplay} (Action/Racing);
\textbf{Head 2} shifts to \textit{Genre Exploration}, capturing Adventure/Sports titles (e.g., \textit{Lego Indiana Jones});
\textbf{Head 3} identifies \textit{Collectibles}, uniquely attending to merchandise like \textit{Mario Amiibo};
and \textbf{Head 4} captures \textit{Utilitarian Needs}, focusing on accessories like \textit{Controller Attachments}.
This demonstrates that FLR-GRPO successfully allocates dedicated heads to capture complementary needs (software vs. hardware) absent in the baseline.

\textbf{Quantitative Disentanglement.}
We quantify this improvement using a Disentanglement Score: $DS = (1 - S_{avg}) \times A_{max}$, combining the distinctiveness between heads (inverse cosine similarity $S_{avg}$) and their focus sharpness ($A_{max}$). FLR-GRPO improves the $DS$ by \textbf{80.1\%}, confirming that the model optimizes its "attention budget" to cover multi-faceted user interests—ensuring recommendations cover both core gaming experiences and necessary accessories.

\subsection{Performance Over Popular And Unpopular Items (RQ5)}
For latent reasoning, it is difficult to directly demonstrate why it leads to performance improvements, as can be done with explicit CoT. Therefore, we consider providing indirect evidence. Intuitively, reasoning should be more beneficial for the more challenging aspects of recommendation, where it is expected to deliver greater gains. Long-tail (i.e., unpopular) items are typically more difficult to recommend accurately, while existing methods already perform well on popular items. To test this hypothesis, we divide items into two groups based on their interaction frequency in the training set: the top 20\% of items are categorized as the popular group, while the remaining 80\% are treated as the unpopular group.

\begin{figure}[ht]
    \centering
    {
        \includegraphics[width=0.95\linewidth]{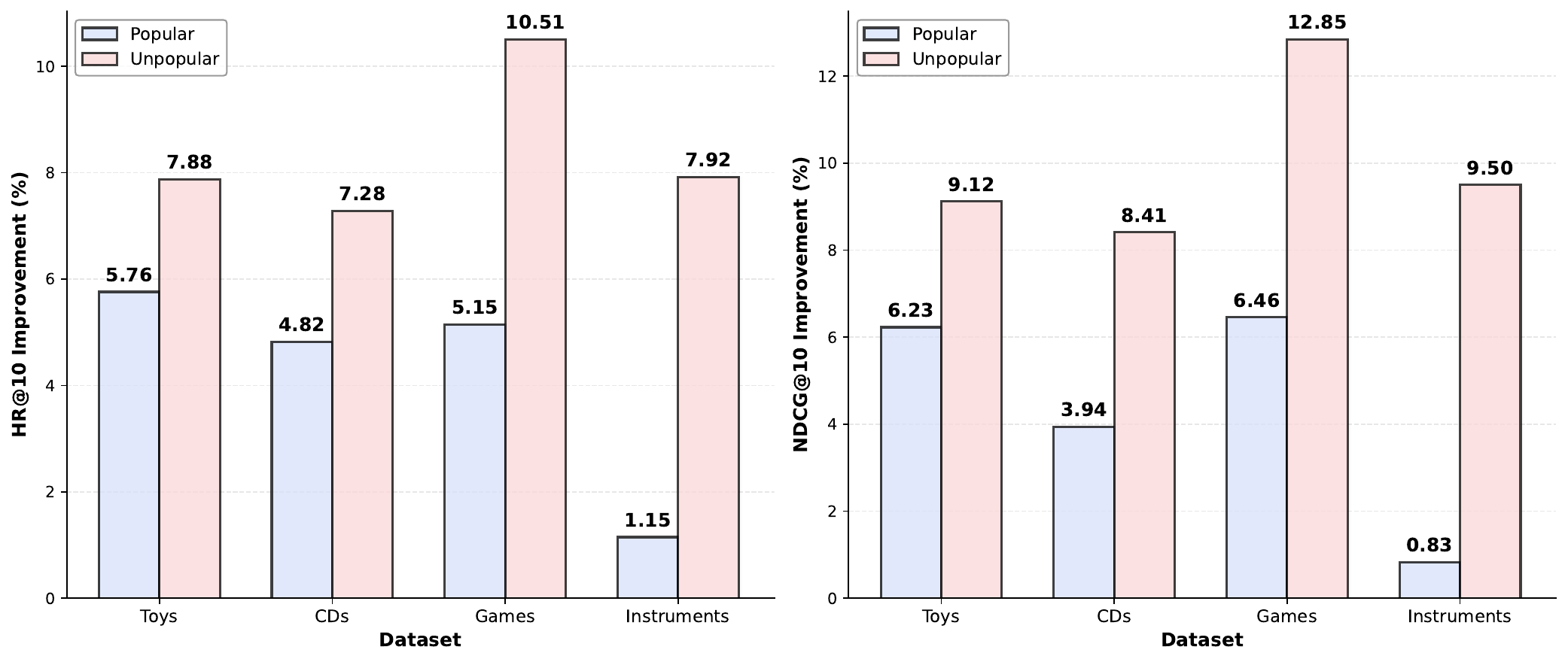}
    }
    \vspace{-1em}
    \caption{Performance improvement of FLR over LatentR\textsuperscript{3} on both popular and unpopular items.}
    \label{fig:popular_unpopular}
\end{figure}

As shown in Figure \ref{fig:popular_unpopular}, FLR achieves consistent performance gains over LatentR\textsuperscript{3} across both groups, but the magnitude of improvement is significantly more pronounced for unpopular items. 
Specifically, on the \textit{Games} dataset, the relative improvement in NDCG@10 is 12.85\% for unpopular items, compared to 6.46\% for popular items. A more pronounced disparity is observed on the \textit{Instruments} dataset, where unpopular items show a 9.50\% improvement in NDCG@10, whereas the gain for popular items is 0.83\%. Consistent patterns are found in HR@10 metrics; for instance, unpopular items in the \textit{Games} and \textit{Instruments} datasets achieve improvements of 10.51\% and 7.92\%, respectively, surpassing the gains seen in the popular group (5.15\% and 1.15\%).
This empirical evidence strongly validates our hypothesis: while popular items already benefit from abundant interaction data, making it harder to achieve large marginal gains, unpopular items suffer from severe data sparsity. The proposed FLR module effectively mitigates this issue by leveraging latent reasoning to infer user intent from semantic clues rather than relying solely on collaborative signals, thereby delivering substantial improvements in the long-tail distribution.

\subsection{Sensitivity to the Number of Latent Reasoning Factors (RQ6)}

\begin{figure}[t]
    \centering
    {
        \includegraphics[width=0.95\linewidth]
        {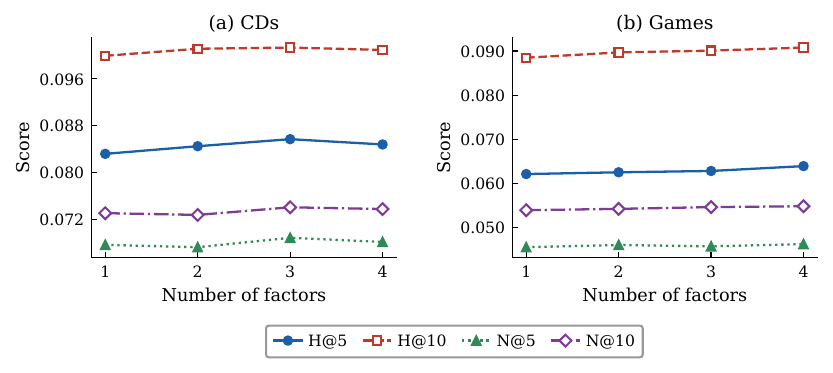}
    }
    \vspace{-1em}
    \caption{Performance of FLR w.r.t. the number of latent reasoning factors.}
    \label{fig:factor_analysis}
\end{figure}

Figure~\ref{fig:factor_analysis} illustrates the performance of FLR 
under varying numbers of latent reasoning factors ($K \in \{1,2,3,4\}$) 
on the CDs and Games datasets. Two key observations emerge.

\textbf{Multi-factor reasoning consistently outperforms single-factor.}
Across both datasets, $K{=}1$ yields the lowest performance on all metrics, 
confirming that a single latent factor is insufficient to capture the 
diversity of user intent. This provides direct evidence for the necessity 
of the multi-head factorization design in FLR.

\textbf{The optimal $K$ is domain-dependent but stable.}
On CDs, performance peaks at $K{=}3$ (H@5$=$0.0857, N@5$=$0.0689) and 
marginally declines at $K{=}4$, suggesting that three orthogonal factors 
adequately span the intent space of this domain, while additional factors 
introduce redundancy. In contrast, Games exhibits a monotonically 
increasing trend up to $K{=}4$ (H@5$=$0.0639, N@5$=$0.0462), reflecting 
the greater diversity of user interests in the gaming domain, which 
benefits from a richer factor pool. Notably, the overall performance 
variation across different values of $K$ remains modest, indicating that 
FLR is robust to the exact factor configuration and does not require 
fine-grained tuning. 

\subsection{Inference Cost Comparison (RQ7)}

Compared to non-reasoning LLM baselines, our method introduces the extra cost of generating latent reasoning tokens. However, since only a few tokens are used (one by default), the added cost is minimal. In contrast, explicit reasoning methods often require a large and uncontrollable number of tokens, leading to significant overhead. We randomly selected 100 samples from each of the four datasets. Inference was performed using a single A100 GPU, with batch size set to 4 and beam size set to 10 (the CoT method set to 1). 

\begin{figure}[t]
    \centering
    {
        \includegraphics[width=0.75\linewidth]{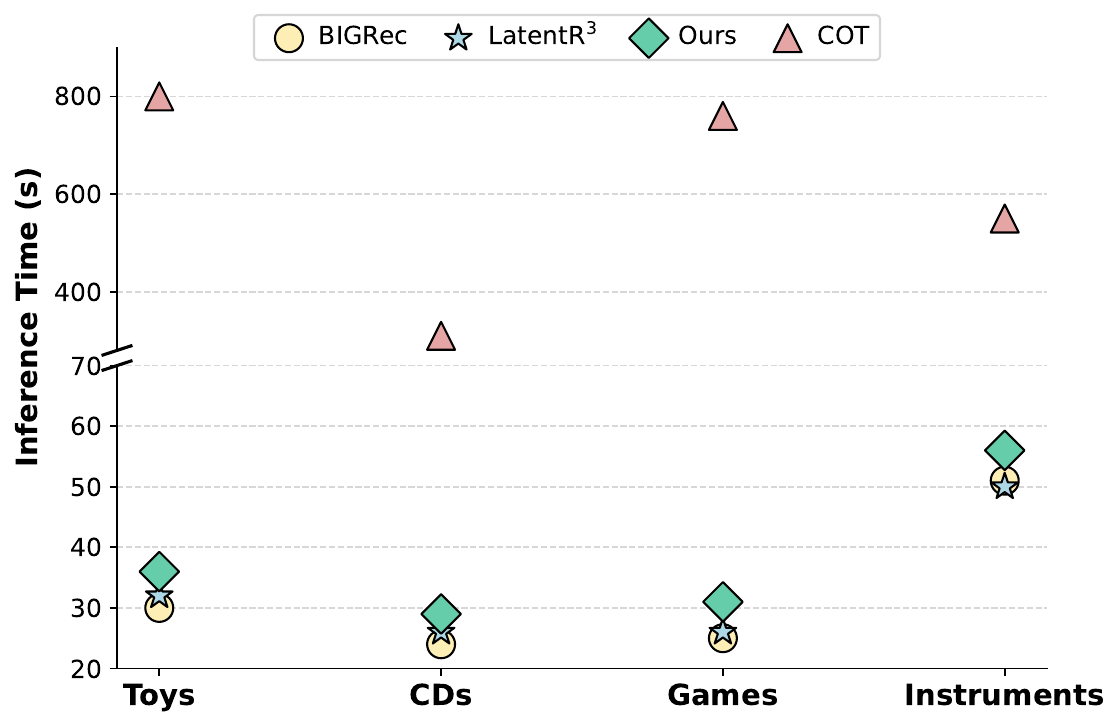}
    }
    \vspace{-1em}
    \caption{Inference time comparison across non-reasoning (BIGRec), LatentR\textsuperscript{3}, FLR (Ours), and explicit CoT methods (CoT).}
    \label{fig:inference_cost}
\end{figure}

Each measurement was taken three times and we report the average values, as shown in Figure \ref{fig:inference_cost}. Because only a single token is added, and the length of item titles inherently varies, our inference time is almost identical to that of non-reasoning methods. However, explicit CoT methods are far more expensive as shown in Figure \ref{fig:inference_cost}.

\section{Conclusion}

We propose Factorized Latent Reasoning (FLR), a parameter-efficient latent reasoning framework for LLM-based recommendation. FLR represents user intent with multiple disentangled preference factors and stabilizes factor learning through orthogonality, diversity, and sparsity constraints. We further introduce a GRPO-based tuning strategy that explores the latent space and optimizes a hybrid reward without token-level reasoning overhead. Experiments on multiple Amazon domains show consistent improvements in top-$K$ ranking over strong baselines with efficient inference.

\section{Acknowledgement}

This work is supported by the Beijing Nova Program.

\bibliographystyle{ACM-Reference-Format}
\bibliography{sample-base}

\end{document}